\documentclass[aps,prl,twocolumn,showpacs,psfig,superscriptaddress,longbibliography]{revtex4-1}
\usepackage{amsfonts}
\usepackage{mathrsfs}
\usepackage{amsmath}
\usepackage{color}
\usepackage{natbib}
\usepackage{graphicx}
\usepackage{bm}
\usepackage{amssymb}
\usepackage{xspace}
\usepackage{epstopdf}
\usepackage{dcolumn}
\usepackage{longtable}
\usepackage{multirow}
\usepackage[colorlinks=true, letterpaper=true, pdfstartview=FitV, linkcolor=blue, citecolor=blue, urlcolor=blue]{hyperref}
\usepackage{float}

\makeatletter

\newcommand{\Rmnum}[1]{\expandafter\@slowromancap\romannumeral #1@}
\makeatother

\begin{document}

\title{Two-dimensional Weyl Half Semimetal and Tunable Quantum Anomalous Hall Effect in Monolayer PtCl$_{3}$}

 \author{Jing-Yang You}
 \affiliation{School of Physical Sciences, University of Chinese Academy of Sciences, Beijing 100049, China}

 \author{Cong Chen}
 \affiliation{Department of Physics, Key Laboratory of Micro-nano Measurement-Manipulation and Physics (Ministry of Education), Beihang University, Beijing 100191, China}
 \affiliation{Research Laboratory for Quantum Materials, Singapore University of Technology and Design, Singapore 487372, Singapore}

 \author{Zhen Zhang}
 \affiliation{School of Physical Sciences, University of Chinese Academy of Sciences, Beijing 100049, China}

 \author {Xian-Lei Sheng}
 \email{xlsheng@buaa.edu.cn}
 \affiliation{Department of Physics, Key Laboratory of Micro-nano Measurement-Manipulation and Physics (Ministry of Education), Beihang University, Beijing 100191, China}
 \affiliation{Research Laboratory for Quantum Materials, Singapore University of Technology and Design, Singapore 487372, Singapore}

 \author{Shengyuan A. Yang}
 \email{shengyuan\_yang@sutd.edu.sg}
 \affiliation{Research Laboratory for Quantum Materials, Singapore University of Technology and Design, Singapore 487372, Singapore}

 \author{Gang Su}
 \email{gsu@ucas.ac.cn}
 \affiliation{School of Physical Sciences, University of Chinese Academy of Sciences, Beijing 100049, China}
 \affiliation{Kavli Institute for Theoretical Sciences, and CAS Center for Excellence in Topological Quantum Computation, University of Chinese Academy of Sciences, Beijing 100190, China}

\begin{abstract}
We propose a new topological quantum state of matter---the two-dimensional (2D) Weyl half semimetal (WHS), which features 2D Weyl points at Fermi level belonging to a single spin channel, such that the low-energy electrons are described by fully spin-polarized 2D Weyl fermions. We predict its realization in the ground state of monolayer PtCl$_3$. We show that the material is a half metal with an in-plane magnetization, and its Fermi surface consists of a pair of fully spin-polarized Weyl points protected by a mirror symmetry, which are robust against spin-orbit coupling. Remarkably, we show that the WHS state is a critical state at the topological phase transition between two quantum anomalous Hall insulator phases with opposite Chern numbers, such that a switching between quantum anomalous Hall states can be readily achieved by rotating the magnetization direction. Our findings demonstrate that WHS
offers new opportunity to control the chiral edge channels, which will be useful for designing new topological electronic devices.
\end{abstract}
\maketitle


{\color{blue}{\em Introduction.}}---Weyl semimetals have been attracting extensive attention in recent research~\cite{Murakami2007,Balents2011,Wan2011,Chiu2016,Burkov2016,Yang2016,Dai2016,Bansil2016,Vishwanath2018}. In a Weyl semimetal, the conduction and valence bands cross linearly at isolated twofold degenerate nodal points in the Brillouin zone (BZ), such that the low-energy electrons resemble the relativistic Weyl fermions. Thus, many intriguing phenomena in relativity and high-energy physics can be explored in condensed matter experiments~\cite{Nielsen1983,Son2013,Guan2017}. In order to achieve the Weyl point, it is necessary to break the inversion ($\mathcal{P}$) or the time reversal ($\mathcal{T}$) symmetry to remove the spin degeneracy of the bands.
So far, most Weyl semimetals are realized in crystals with broken $\mathcal{P}$, while the candidates with broken $\mathcal{T}$, i.e., the magnetic Weyl semimetals, are much less~\cite{Wan2011,XuG2011PRL,Wang2016,Kubler2016,XuQN2018,Morali2019}.
Moreover, the studies are mainly for three dimensional (3D) systems. A Weyl point in 3D has a topological protection, characterized by the Chern number defined on a surface enclosing the point. In comparison, a Weyl point in 2D must require additional symmetry protection~\cite{Zhao-Wang-Classification}. Such reduction of protection means that the Weyl semimetal phase in 2D is less robust than its 3D counterpart. On the other hand, however, it also leads to the opportunity to more easily manipulate the topological phase transitions in 2D, especially the interplay between magnetism and band topology if the Weyl phase is realized in a magnetic state.

In this work, we propose a new topological state in 2D---the 2D Weyl half semimetal (WHS), which is both a half metal and a semimetal, with fully spin-polarized Weyl points at Fermi level formed in a single spin channel. Consequently, the low-energy electrons are \emph{fully} spin-polarized 2D Weyl fermions. We predict the realization of this novel phase in monolayer PtCl$_3$. Based on first-principles calculations, we show that the ground state of monolayer PtCl$_3$ is a 2D WHS with an in-plane magnetization, which preserves a vertical mirror plane. A pair of 2D Weyl points are protected by the mirror symmetry and are robust even under spin-orbit coupling (SOC). Furthermore, we find that the 2D WHS state represents a critical point between two quantum anomalous Hall (QAH) insulator phases with opposite Chern numbers $\pm 1$. By breaking the mirror, e.g., by rotating the magnetization vector, one can readily control the realization of QAH phases and the propagating direction of the chiral edge channels. Our findings not only reveal a new state of matter, but also offer promising material platforms for novel topological spintronics applications.

\begin{figure}[!hbt]
  \centering
  \includegraphics[scale=0.8,angle=0]{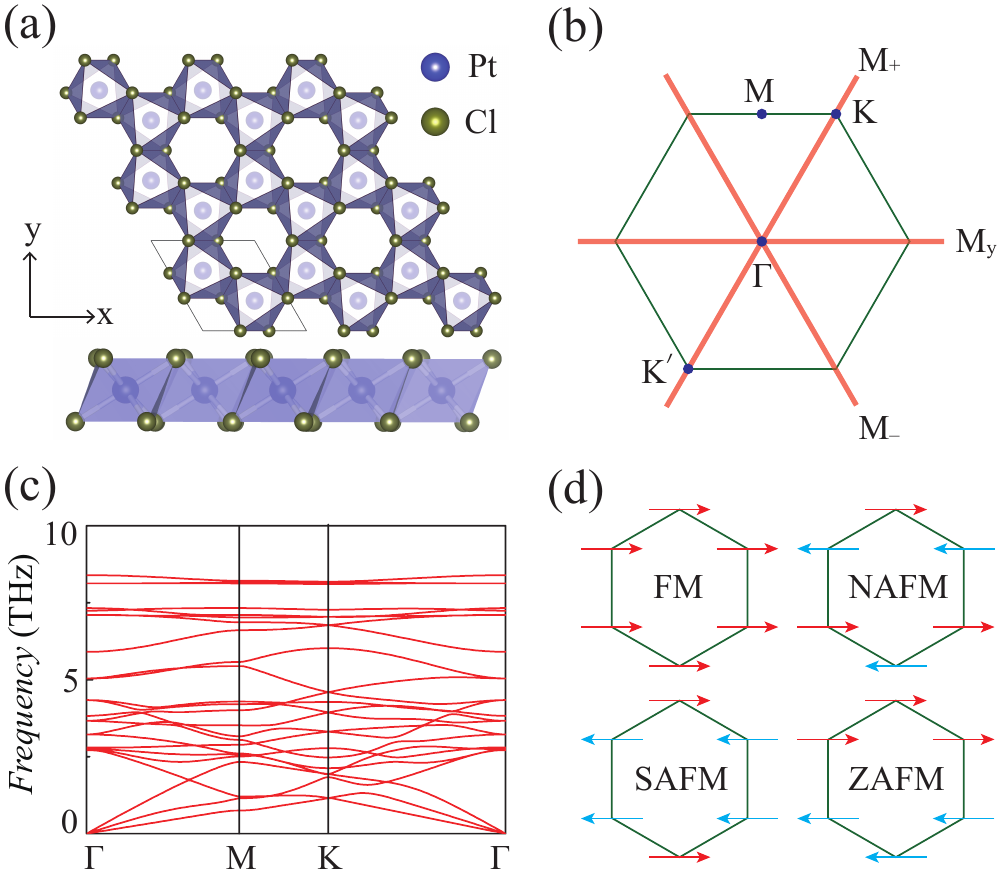}\\
  \caption{(a) Top and side view of monolayer PtCl$_{3}$, with edge sharing PtCl$_6$ octahedron forming a honeycomb lattice.  (b) First Brillouin zone for monolayer PtCl$_{3}$ with high symmetry points labeled. We also mark the orientation of the three vertical mirror planes for the lattice structure (red lines). (c) Phonon spectrum for monolayer PtCl$_{3}$. (d) Possible magnetic configurations considered: ferromagnet (FM), N\'{e}el antiferromagnet (NAFM), stripe AFM (SAFM), and zigzag AFM (ZAFM).  The magnetic moments are on the Pt sites forming a honeycomb lattice.
  }\label{fig1}
\end{figure}

{\color{blue}{\em Computational method.}}---Our first-principles calculations were based on the density-functional theory (DFT) as implemented in the Vienna \textit{ab initio} simulation package (VASP)~\cite{Kresse1994,Kresse1996}, using the projector augmented wave method~\cite{PAW}. The generalized gradient approximation with Perdew-Burke-Ernzerhof~\cite{PBE} realization was adopted for the exchange-correlation functional. The plane-wave cutoff energy was set to 520 eV. The Monkhorst-Pack $k$-point mesh~\cite{PhysRevB.13.5188} of size $11\times11\times 1$ was used for the BZ sampling.  To account for the correlation effects for transition metal elements, the DFT$+U$ method~\cite{Anisimov1991,Dudarev1998} was used for calculating the band structures.  The crystal structure was optimized until the forces on the ions were less than 0.01 eV/\AA. The surface spectrum was calculated by using the Wannier functions and the iterative Green's function method~\cite{Marzari1997,Souza2001,Wu2017,Green}.

{\color{blue}{\em Structure and magnetism.}}---Monolayer PtCl$_3$ consists of a Pt atomic layer sandwiched by two Cl atomic layers, where the Pt atoms form a honeycomb lattice and each Pt is surrounded by six Cl atoms forming an octahedral crystal field, as shown in Fig.~\ref{fig1}(a). It takes the same structure as monolayer CrI$_3$~\cite{Huang2017} and RuCl$_3$~\cite{Banerjee2016} that have been shown to be 2D magnetic materials. The point group symmetry is $D_{3d}$, with generators of a rotoreflection $S_6$ and a vertical mirror $\sigma_d$. Combining these two operations leads to another two vertical mirror planes, as illustrated in Fig.~\ref{fig1}(b). The three vertical mirrors play an important role in the discussion of the WHS state below. The optimized lattice constant from our first-principles calculations is 6.428 \AA. To confirm its stability, we calculate the phonon spectrum, which shows no imaginary frequency mode [see Fig.~\ref{fig1}(c)], indicating that monolayer PtCl$_3$ is dynamically stable.

\begin{figure}[t]
  \centering
  \includegraphics[scale=1.0,angle=0]{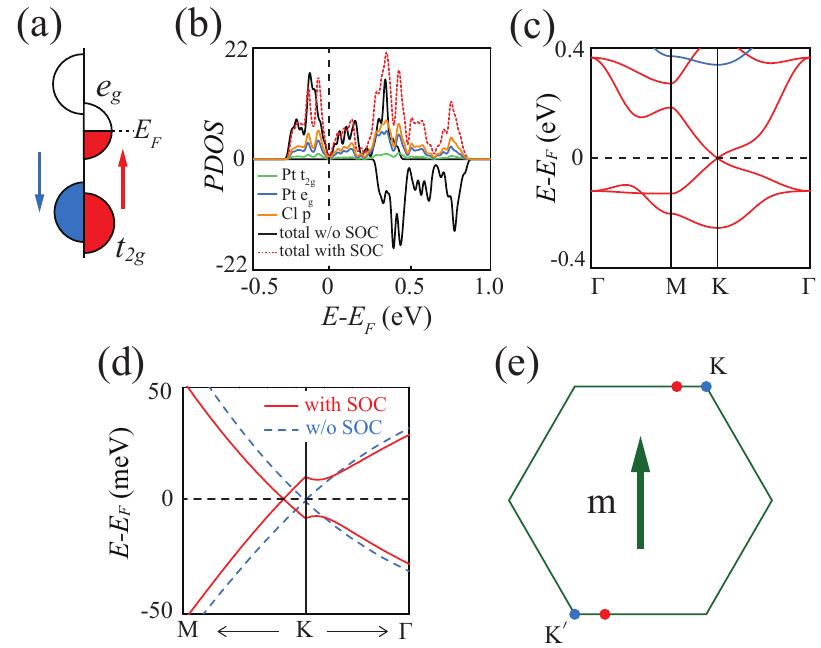}\\
  \caption{(a) Schematic depiction of the orbital splitting in monolayer PtCl$_3$. (b) Spin-resolved partial density of states (PDOS) for monolayer PtCl$_3$ projected on different orbitals. (c) Band structure without spin-orbit coupling (SOC). The red and blue bands are for spin majority (spin-up) and minority (spin-down) channels, respectively. (d) Enlarged view of the band structure around the Weyl point. The red solid (blue dashed) lines are for the bands with (without) SOC. (e) Two Weyl points are located at $K/K'$ points without SOC (blue points), and they are shifted along $x$ direction on the mirror-invariant line after considering SOC (red points).}\label{fig2}
\end{figure}

\begin{table}[t]
	\caption{The total energy $E_\mathrm{tot}$ per unit cell (in meV, relative to $E_\mathrm{tot}$ of the FM$^y$ ground state) as well as spin $\langle S \rangle$ and orbital $\langle O \rangle$ moments (in $\mu_{B}$) for several magnetic configurations calculated by GGA+SOC+$U$ method. The superscript in each configuration indicates the magnetic polarization direction.  }\label{tab:magnet}\label{tab1}
	\begin{tabular}{cccccccc}
		\hline
		                     & FM$^y$ & NAFM$^y$ & SAFM$^y$ & ZAFM$^y$ &FM$^z$ &FM$^x$ &PM\\
		\hline \hline
		$E_\mathrm{tot}$     &0.00    &316.61    &233.84    &96.19     &5.29    &0.67  &331.07\\
		$\langle S \rangle$  &0.76    &0.80      &0.78      &0.77      &0.76    &0.76  &0\\
		$\langle O \rangle$  &0.19    &0.23      &0.20      &0.20      &0.25    &0.19  &0\\
		\hline
	\end{tabular}
\end{table}


Pt is a transition metal element with partially filled $d$ shell, which may give rise to magnetism. Indeed, our first-principles calculations show that monolayer PtCl$_3$ favors a ferromagnetic (FM) ground state than the antiferromagnetic (AFM) or the paramagnetic (PM) states [see Fig.~\ref{fig1}(d) and Table~\ref{tab1}]. Furthermore, the FM state is found to be a half metal, i.e., with a single spin channel present at the Fermi level, as can be observed from the projected density of states (DOS) in Fig.~\ref{fig2}(b) and the band structure in Fig.~\ref{fig2}(c).

To understand this, we note that under the octahedral crystal field, Pt-$5d$ orbitals are split into $t_{2g}$ and $e_g$ groups, with the latter energetically higher. For Pt$^{3+}$ with seven valence electrons, Pt-$t_{2g}$ orbitals will be fully-filled. Since the crystal field in PtCl$_3$ is stronger than exchange field, the fully-filled $t_{2g}$ orbitals are away from the Fermi level. On the other hand, Pt-$e_g$ orbitals are filled by one electron, hence are fully spin-polarized. Because there are two Pt atoms in the primitive cell, the bands dominated by $e_g$ orbitals are half-filled for one spin channel and empty for the other, as schematically depicted in Fig.~\ref{fig2}(a). The bands around Fermi level are completely from the spin-up subband of $e_g$ orbitals, therefore making it a half metal with 100$\%$ spin polarization.

Next, we shall pin down the magnetization direction for the FM ground state. We compare the energies by scanning the magnetization direction $\bm m$ (with SOC included), and find that: (i) in-plane directions are energetically preferred over the out-of-plane ones; (ii) among the in-plane ones, the directions perpendicular to the vertical mirrors (i.e., the armchair direction for the Pt honeycomb lattice) have the lowest energy (see Table~\ref{tab1}).
It follows that the magnetic interaction around the ground state configuration may be approximately described by the following spin Hamiltonian
\begin{equation}
\begin{split}
H&=-\sum_{\langle i,j\rangle} J(S_{i}^{x}S_{j}^{x}+S_{i}^{y}S_{j}^{y})-\sum_{i} D(S_{i}^{y})^{2},
\end{split}
\end{equation}
where $S^{x,y}$ is the spin operator, $\langle i,j\rangle$ denotes the summation over nearest neighboring sites, $J$ and $D$ denote the strengths for exchange interaction and anisotropy, respectively. The values of $J$ and $D$ can be extracted from the first-principles calculations.
Approximating the model as an anisotropic 2D $XY$ ferromagnet, the Curie temperature for the FM state can be estimated~\cite{2DXY,Costa1996,Ma1997} as $T_C\approx 200 $ K.

{\color{blue}{\em 2D Weyl half semimetal.}}---In the band structure plot in Fig.~\ref{fig2}(c), one notices a remarkable feature: the conduction and valence bands form a linear crossing point at the Fermi level. Since the two crossing bands are fully spin polarized (spin-up), the crossing point is twofold degenerate and represents a 2D Weyl point. Thus, the ground state for PtCl$_3$ is a 2D WHS, with the low-energy electrons being 100\% spin-polarized 2D Weyl fermions.

In the absence of SOC, a pair of Weyl points are located at the $K$ and $K'$ points of the BZ, similar to graphene, but they are formed by a single spin species. Without SOC, the spin and the orbital part of the electronic wave function are decoupled, and hence all crystalline symmetries are preserved for each spin channel separately as for spinless particles. To characterize the low-energy band structure, we construct a $k\cdot p$ effective model expanded around the $K/K'$ point. It is subjected to the $C_{3v}$ little group at $K$ ($K'$), with two generators $C_{3z}$ and $M_y$. The effective Hamiltonian must satisfy
\begin{equation}\label{eqC}
C_{3z}\mathcal{H}_0(q_+,q_-)C^{-1}_{3z} = \mathcal{H}_0(q_{+}e^{i2\pi/3}, q_{-}e^{-i2\pi/3}),
\end{equation}
\begin{equation}\label{eqM}
M_y\mathcal{H}_0(q_x,q_y)M^{-1}_y = \mathcal{H}_0(q_x, -q_y),
\end{equation}
where $\bm q$ is measured from $K$, and $q_\pm=q_x\pm iq_y$. And the two Weyl points are related by inversion.
In the basis of the 2D irreducible representation $E$ for $C_{3v}$, we find that to linear order in $q$, the effective model takes the form of the 2D Weyl model
\begin{equation}\label{eqHeff}
\mathcal{H}_0(\bm{q})= v_F(\tau q_x\sigma_x +q_y\sigma_y),
\end{equation}
where $v_F$ is the Fermi velocity, $\tau=\pm$ for the $K/K'$ point,  and $\sigma_i$'s are the Pauli matrices acting in the space of the two basis states. Thus, the low-energy electrons indeed resemble 2D Weyl fermions. It is worth noting that despite the similarity to the low-energy model for graphene~\cite{CastroNeto2009}, the model basis and hence the described fermions here are fully spin polarized.

As we have mentioned, the inclusion of SOC pins the ground state magnetization perpendicular to one of the vertical mirrors (taken to be $M_y$ here). It follows that the $C_{3z}$ symmetry is broken but the $M_y$ symmetry is still preserved. The preserved $M_y$ dictates that the spin-up and spin-down bands are still fully spin polarized (along $y$) \emph{without} hybridization by SOC. Remarkably, one finds that the two Weyl points are maintained, only their locations slightly shifted from $K$ and $K'$ to some nearby points on the path $K$-$M$ and $K'$-$M$ which are invariant under the remaining mirror [see Fig.~\ref{fig2}(d) and \ref{fig2}(e)]. The Weyl points are still protected, since the two crossing bands have opposite $M_y$ eigenvalues. On the level of the effective model, to leading order in $k$, SOC introduces the following term
\begin{equation}
\mathcal{H}_\text{SOC}=\eta\sigma_x
\end{equation}
for both $K/K'$ points. As a result, the original Weyl point at $K/K'$ is shifted by $\mp \eta/v_F$ along the $x$ direction (i.e., on the mirror-invariant line) but does not open a gap. This is consistent with the first-principles calculation result in Fig.~\ref{fig2}(d). We mention that it is quite rare to have robust 2D Weyl point under SOC~\cite{WuWK2019}. To our knowledge, this is the first time to find such Weyl point in a magnetic state.

The above discussion demonstrates that the ground state of monolayer PtCl$_3$ is indeed a WHS with a pair of fully spin-polarized 2D Weyl points, and this state is robust under SOC. It is in contrast to the Dirac points in graphene, which are unpolarized and are removed when SOC is turned on.
Below, we shall show that the WHS state represents a critical point between two QAH insulator phases with Chern numbers $\mathcal{C}=\pm 1$. This in turn requires that the WHS state must be gapless, since it is located at a topological phase transition.

\begin{figure}[tbp]
  \centering
  \includegraphics[scale=0.95,angle=0]{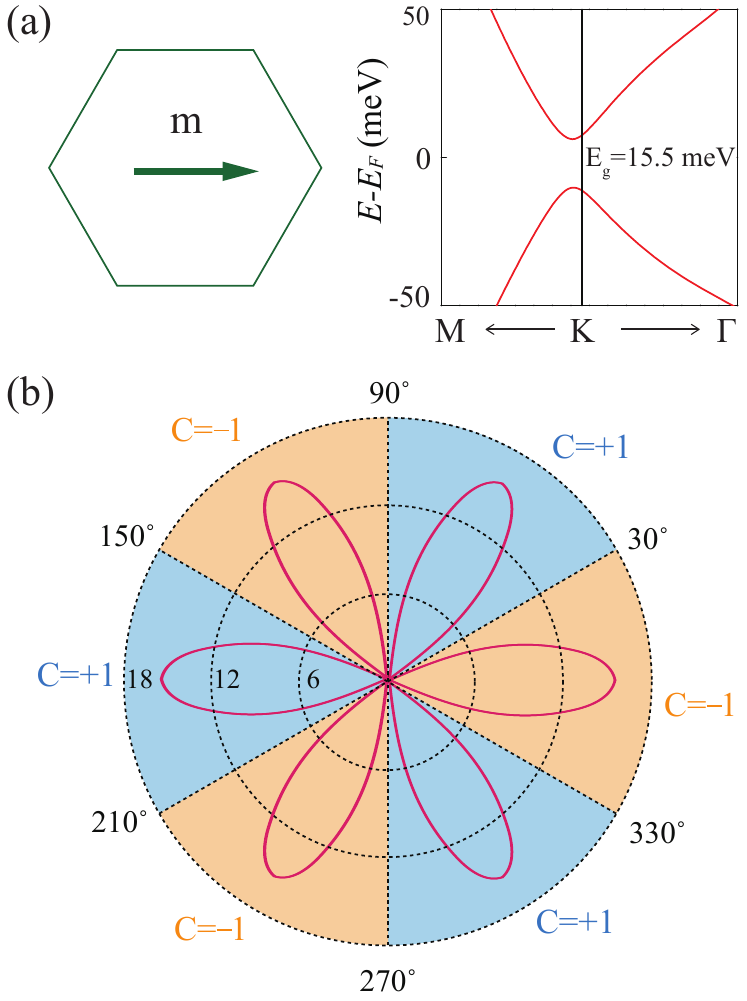}\\
  \caption{ (a) A gap is opened at the original Weyl point when the magnetization is along the zigzag direction. (b) The flower-like curve (red line) shows the band gap as a function of the azimuthal angle $\phi$ for the magnetization direction, where polar radius indicates the gap value (in meV). The blue (orange) color indicates the regions with Chern number $\mathcal{C}=+1$ ($-1$). }\label{fig3}
\end{figure}

{\color{blue}{\em Tuning QAH phases.}}---Because the Weyl points are protected by $M_y$, breaking $M_y$ will generally remove the Weyl points and open an energy gap. For example, the result for $\bm m$ along the $x$ direction (zigzag direction for the Pt honeycomb lattice) is shown in Fig.~\ref{fig3}(a). Clearly, a finite band gap $\sim 15.5$ meV is opened at the original Weyl point. In the effective model, this adds a mass term $\mathcal{H}_\Delta=\frac{\Delta}{2}\sigma_z$ with $|\Delta|$ the gap size, such that the effective model becomes
\begin{equation}
  \mathcal{H}=\mathcal{H}_0+\mathcal{H}_\text{SOC}+\mathcal{H}_\Delta.
\end{equation}

It is known that the gap opening at a 2D Weyl point would induce a finite Berry curvature $\Omega(\bm q)=-2\text{Im}\langle \partial_{q_x} u_v|\partial_{q_y} u_v\rangle$, where $|u_v\rangle$ is the eigenstate of the valence band.
The integral of Berry curvature in a region around the Weyl point gives a valley topological charge of $\pm 1/2$~\cite{YaoW2009,PanH2015}, with the sign determined by  $\text{sgn}(\Delta)$. For monolayer PtCl$_3$, $\mathcal{T}$ is broken, and the two Weyl points are related by $\mathcal{P}$. Because the Berry curvature is an even function under $\mathcal{P}$, the valley topological charge for the two points after gap opening must be the same, and therefore a finite Chern number $\mathcal{C}=\int_\text{BZ}\Omega(\bm k) d\bm k=\pm 1$ must be resulted. This indicates that gapping the WHS state will generate a QAH insulator phase.

The analysis above is confirmed by the first-principles calculations. In Fig.~\ref{fig3}(b), we plot the band gap and the Chern number as functions of angle $\phi$, which is the azimuthal angle for the magnetization vector $\bm m$, assuming $\bm m$ is rotated in-plane. One observes that the gap vanishes
at $\phi=\pm\frac{\pi}{6}$, $\pm\frac{\pi}{2}$, and $\pm\frac{5\pi}{6}$, at which one of the mirror planes is preserved, and the state corresponds to  a WHS. In regions between these values, the gap becomes nonzero, and the Chern number takes values alternating between $+1$ and $-1$.
Since the finite band gap and the (quantized) Hall conductivity $\sigma_{xy}=\frac{e^2}{h}\mathcal{C}$ are tied together in the current case, the gap closing in the WHS state can also be understood as a general symmetry requirement. It was shown by Liu \emph{et al.}~\cite{LiuCX2013} that to maintain the invariance of the Hall response equation $j_x=\sigma_{xy}E_y$ for a nonzero $\sigma_{xy}$, all the vertical mirrors must be broken. Thus, at the special values of $\phi$ that preserve a mirror, $\sigma_{xy}$ hence the gap must be zero.

\begin{figure}[tbp]
  \centering
  \includegraphics[scale=0.95,angle=0]{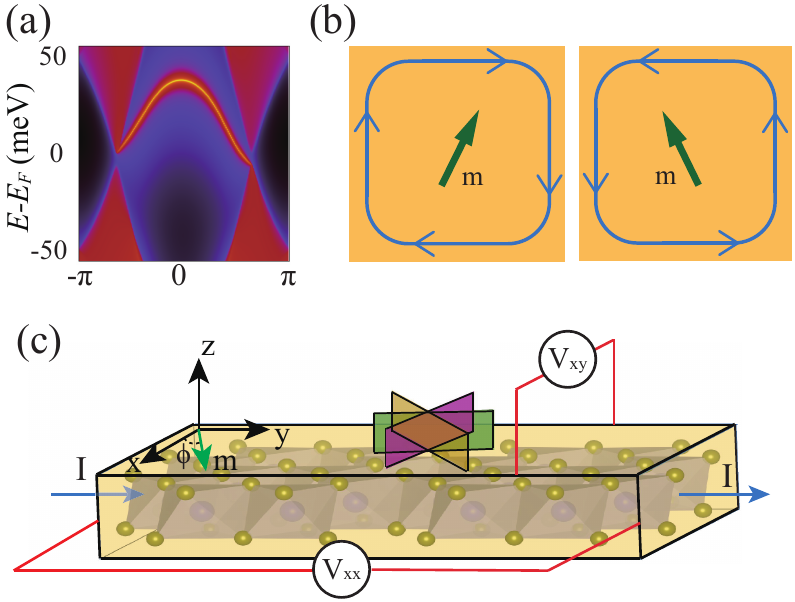}\\
  \caption{(a) The edge spectrum corresponding to the case in Fig.~\ref{fig3}(a), showing the existence of gapless chiral edge states. (b) Schematic top views of a finite size sample. By tuning the magnetization direction to regions with opposite Chern numbers, one can switch the propagation direction of the chiral edge channel. This can be probed by the standard transport measurement setup as in (c).
  }\label{fig4}
\end{figure}

The hallmark of the QAH phase is the existence of chiral edge states, i.e., gapless channels at the edge propagating unidirectionally. Figure~\ref{fig4}(a) shows the edge spectrum obtained from first-principles calculations, which confirms the existence of one chiral channel per edge. The chirality of the edge channel is determined by the sign of Chern number. Consequently, by tuning across the topological phase transition at the WHS state, the propagating direction of the edge channel will be reversed [see Fig.~\ref{fig4}(b)]. This can be detected in electrical transport measurement as shown in Fig.~\ref{fig4}(c).

{\color{blue}{\em Discussion.}}---We have revealed a new topological quantum state---the 2D WHS, and demonstrated its realization in monolayer PtCl$_3$. Since it is a critical state at the topological phase transition between two QAH phases, it offers great advantage to control the QAH phases. In experiment, the switching can be readily achieved by applying an in-plane magnetic field to rotate the magnetization vector. By switching between $\mathcal{C}=+1$ and  $-1$ states, one changes the propagation direction of the QAH chiral edge channel. This may offer a new mechanism for designing novel topological electronic devices.

The electronic correlation effect could be important for transition metal compounds, although it is typically weak for $5d$ elements like Pt. Here, we test the effect of correlation via the DFT$+U$ approach~\cite{Anisimov1991,Dudarev1998}. We find that the results are qualitatively unchanged for $U$ values up to 2 eV, and only for very large $U$ ($>2.8$ eV), the system can be transformed into a Mott insulator. Since typical $U$ value for $5d$ elements is less than 1.5 eV, the results presented here should be robust.

Finally, we mention that since the WHS state here is protected by the mirror symmetry, it is robust under biaxial strain or uniaxial strains along the high symmetry directions (zigzag or armchair), which preserve the mirror. For more general strains (like shear strain), the WHS would transform into the QAH phase. Strains can further be used to tune the gap of the QAH state. For example, the band gap for the case in Fig.~\ref{fig3}(a) can be increased to $\sim20$ meV under a biaxial 5\% compressive  strain.

\begin{acknowledgements}
We thank D. L. Deng for helpful discussion. This work is supported in part by the National Key R$\&$D Program of China (Grant No. 2018FYA0305800), the Strategic Priority Research Program of CAS (Grant Nos. XDB28000000, XBD07010100), the NSFC (Grant No. 11834014, 14474279, 11504013), Beijing Municipal Science and Technology Commission (Grant No. Z118100004218001), and the Singapore Ministry of Education AcRF Tier 2 (MOE2015-T2-2-144).
\end{acknowledgements}

\ \
\par
\ \
\begin{appendix}
\renewcommand{\theequation}{A\arabic{equation}}
\setcounter{equation}{0}
\renewcommand{\thefigure}{A\arabic{figure}}
\setcounter{figure}{0}
\renewcommand{\thetable}{A\arabic{table}}
\setcounter{table}{0}


\end{appendix}

\bibliographystyle{apsrev4-1}
\bibliography{WHS}

\end{document}